\RequirePackage{lineno}
\documentclass[superscriptaddress,amsmath,amssymb,nofootinbib]{revtex4}

\usepackage{graphicx}
\usepackage{dcolumn}
\usepackage{bm}
\usepackage{ulem}
\usepackage{xfrac}

\usepackage[printwatermark]{xwatermark}
\usepackage{xcolor}

\usepackage{graphicx,amsmath,amssymb}
\usepackage{epstopdf}
\usepackage{lineno}
\usepackage[colorlinks=true, urlcolor=blue, linkcolor=blue, citecolor=blue]{hyperref}

\usepackage{lipsum}
\usepackage{xfrac}

\begin{document}

\title{Relativistic corrections to the rotation curves of disk galaxies}

\author{ Alexandre Deur}

\affiliation{University of Virginia, Charlottesville, VA 22904. USA\\
\email{deurpam@jlab.org} 
}

\begin{abstract}
We present a method to investigate relativistic effects arising from large masses. 
The method is non-perturbative and employs a mean-field approximation and gravitational lensing. 
Using this method and a basic model of disk galaxy, we find that relativistic corrections 
to the rotation curves of disk galaxies are significant at large galactic radii. 
The model predicts a strong correlation between the inferred galactic dark mass and 
the galactic disk thickness, which we verified using two separate sets of observational data.

\end{abstract}

\maketitle

\section{Introduction}

The total mass of a nearby disk galaxy is typically obtained from measuring its rotation curve and deducing from it the mass using
Newton's dynamics. The rationale for this non-relativistic treatment is the small velocity of stars: $v/c \ll 1$ sufficiently far from 
the central galactic black hole. 
However, there are other relativistic corrections beside $(v/c)^n$ terms. Inspecting the
perturbative post-Newtonian~\cite{Einstein:1938yz} Lagrangian for two masses $M_1$ and $M_2$ separated 
by $r$, reveals  terms such as $V_{1pn}=G^2M_{1}M_{2}(M_{1}+M_{2})/2r^2$ 
($G$ is the gravitational constant) that are not suppressed at small $v$. These terms can be non-negligible if
$M_1$ and $M_2$ are large enough, but for galaxies, they happen to be generally small. However,  terms such as $V_{1pn}$  
are {\it{perturbative}} corrections, i.e. they omit non-perturbative dynamics, and we will show here that they indeed fail to provide  
the full relativistic dynamics  associated with large masses.
Regardless of the size of their contribution, terms like $V_{1pn}$ exemplify  
the non-linear nature of General Relativity (GR), which arises from its field self-interaction: 
the gravitational field has an energy and hence gravitates too. 

Field self-interactions are well-known in particle physics: Quantum Chromodynamics (QCD, 
the gauge theory of the strong force between quarks) features color-charged fields that self-interact.
In fact, GR and QCD have similar Lagrangians, including self-interacting terms, as the polynomial form of the 
Einstein-Hilbert Lagrangian shows~\cite{Deur:2009ya, Deur:2016bwq}. 
Field self-interaction in QCD, which causes
quark confinement, occurs even for static sources: numerous 
bound states of heavy quarks (in which quarks have $v\approx 0$) exist~\cite{Tanabashi:2018oca}.
This, and the correspondence of terms in the GR and QCD Lagrangians, 
shows that for bodies massive enough a relativistic treatment is required regardless of their velocity.
 Finally, the measured speeds at the rotation curve plateaus are of several hundreds of km/s, 
 e.g. 300 km/s (or $v/c=0.1\%$) for NGC~2841. They are similar to that of stars which orbit the Galactic
central black hole and which clearly display a relativistic dynamics~\cite{Hees:2017aal}.

These arguments call for an investigation of the importance of relativistic 
dynamics in galaxies. 
The exact resolution of the self-gravitating disk problem within GR being unknown, an approximate method is
required. From experience with QCD, a non-perturbative approach is required to fully account for field self-interaction.
In fact, perturbative QCD underestimates the large distance effect of field self-interaction 
by two orders of magnitude~\cite{pert vs npert QCD}. The identical Lagrangian structure of GR and QCD suggests
similar non-linear issues for galactic dynamics. Furthermore, post-Newtonian calculations show 
that the perturbative series --which may already underestimate field self-interaction--
diverges for masses similar to those of the heaviest known galaxies.
These facts make a perturbative expansion based on terms like $V_{1pn}$ suspicious, and warranty a non-perturbative 
investigation. In Refs.~\cite{Deur:2009ya, Deur:2016bwq}, a non-perturbative 
numerical lattice method was used. Here, we propose to approach the problem with a mean-field
technique combined with gravitational lensing. The advantages 
compared  to the lattice method~\cite{Deur:2009ya, Deur:2016bwq} are:
(1) it is an entirely independent method, thereby providing a thorough check of the lattice result;
(2) it is not restricted to the static limit of the lattice method and can be applied to systems with complex geometries;
(3) it is less CPU-intensive than a lattice calculation, and hence faster;
(4) it clarifies that the effect calculated in Refs.~\cite{Deur:2009ya, Deur:2016bwq} is classical. The lattice approach 
-- an inherently quantum field theory (QFT) technique -- may  
suggest that quantum phenomena are involved. However, the lattice calculations~\cite{Deur:2009ya, Deur:2016bwq} were 
performed in the high-temperature, i.e. classic, limit~\cite{Deur:2016bwq};
(5) the lensing formalism is more familiar to astronomers, in contrast to lattice techniques with its QFT underpinning and terminology. 
 
\section{Mean-field lensing model}
\subsection{Method}
\begin{figure}[tb]
\center
\includegraphics[width=0.45\textwidth]{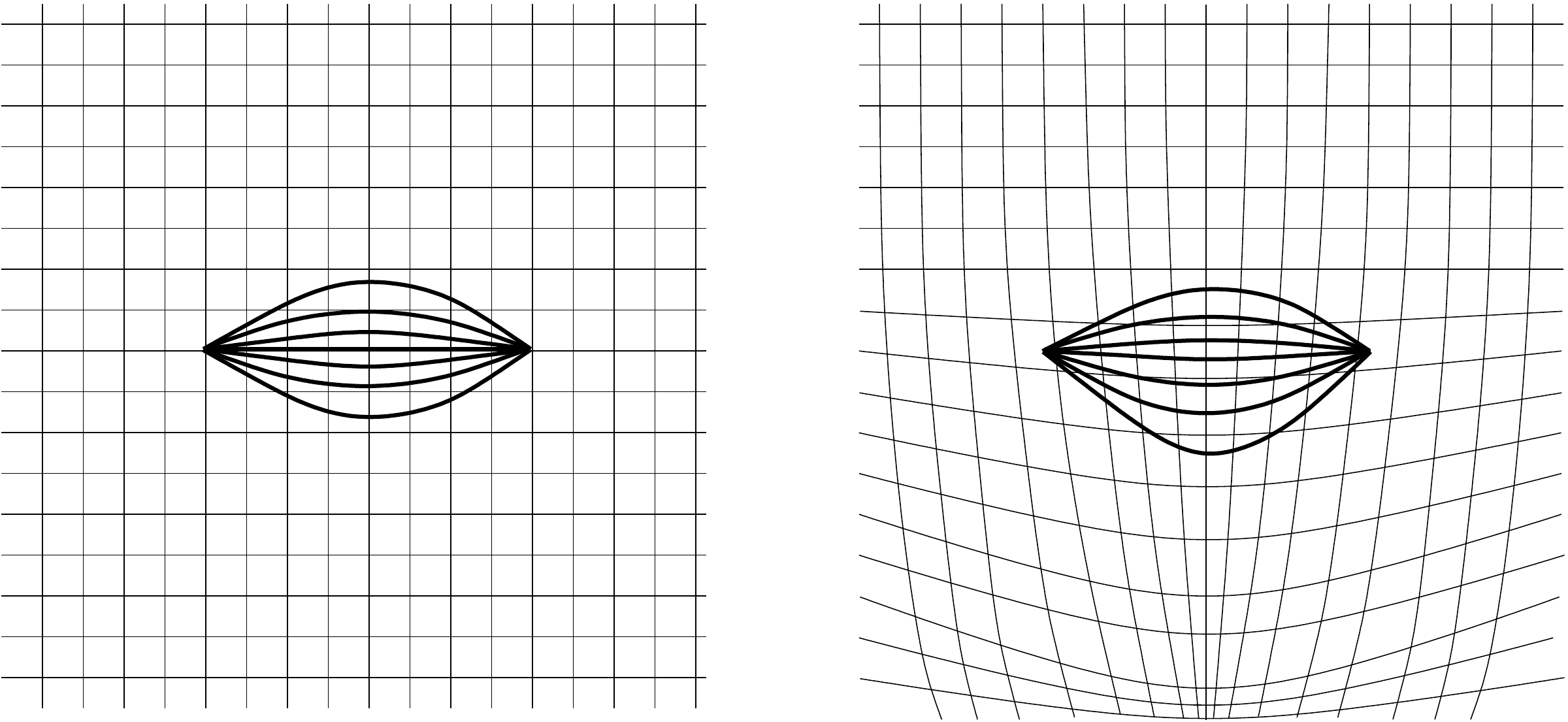} 
\vspace{-0.3cm}
\caption{{\small Illustration of the background field method. The left sketch shows field lines between two masses or charges in 
flat spacetime. The right panel shows the field lines deformation in spacetime curved by a background field. In this article, the mass
deforming spacetime is the total galactic mass and the effect on the field lines connecting two parts of the galaxy is determined by using
the standard gravitational lensing formalism.    
\label{fig:sketch_bkg}}}
\end{figure}

The mean-field technique has been widely employed. In the context of gravitation, it has been used e.g.~to 
calculate the propagation of gravitational waves near large masses~\cite{Landau-Lifshitz2}; 
to solve, by using the gravity/gauge correspondence, the non-perturbative bound-state problem of QCD~\cite{Brodsky:2014yha};
or to derive the background field method introduced to quantize gravity~\cite{DeWitt:1967ub}. 
Hence the mean-field approach is a common and diverse technique. We propose to use it together with the gravitational lensing formalism
to compute the self-interaction of the gravitational fields that generates GR's characteristic non-linearities.
Here, the mean field represents the overall effect of the massive components of the galaxy and is treated
as a spacetime curvature. 
The gravitational interaction between a test particle at a distance $r$ and the 
mass inclosed within $r$, is treated as a traditional force. 
Once this part of the gravitational field is treated as a force rather than geometrically, it can be characterized by field lines. 
The effect of curvature on the traditional force presents in the curved
spacetime accounts for the gravitational field self-interaction. To see this, one can imagine that the 
traditional force acting on the test particle is the electric force rather than gravity. 
Spacetime curvature does affect the electric force~\cite{Smith:1980tv} and it represents the gravitational and electromagnetic fields interacting. 
Thus, for a gravitational force acting on test particle, the effect addressed by the present method is indeed gravity's self-interaction.
This picture of the method is shown in Fig.~\ref{fig:sketch_bkg}.

We will compute here how gravity's field lines are distorted by a mass distribution in the same 
way as light is lensed by such distribution~\cite{field line=trajectories}. 
Then, the field line flux is computed to obtain the gravitational force including its self-interaction effects. 
The usual lensing formalism can be employed because electromagnetic and gravitational field lines are affected 
identically by curved spacetime, that is, in particle language, photons 
and gravitons follow the same null geodesics~\cite{Ragusa:2003kn}.

\subsection{Calculations and model of disk galaxy \label{Calculations and Model of disk galaxy}}

The disk galaxy producing the background field is modeled as an axisymmetric homogeneous 
disk of surface brightness $I(R,z)$ decreasing exponentially with projected radius $R$ and altitude $z$ according to the
characteristic radial scale $h_R$ and scale height $h_z$, respectively. 
While the field lines originate from the whole mass distribution,  considering only  those that 
stem from the galaxy center happens to yield representative results, as it can be 
intuited from the large baryonic matter density around the galactic center and its fast decrease with radius. 
Thus, for simplicity we will first consider only field lines emerging from the galaxy center and
refer the reader to Appendix A for the full calculation accounting for the fluxes originating from the whole mass distribution.
We note that although the simplified and full calculations
happen to yield similar results, the full calculation reveals that the central galactic nucleus plays a critical role. 

\begin{figure}[tb]
\center
\includegraphics[width=0.4\textwidth,height=0.33\textwidth]{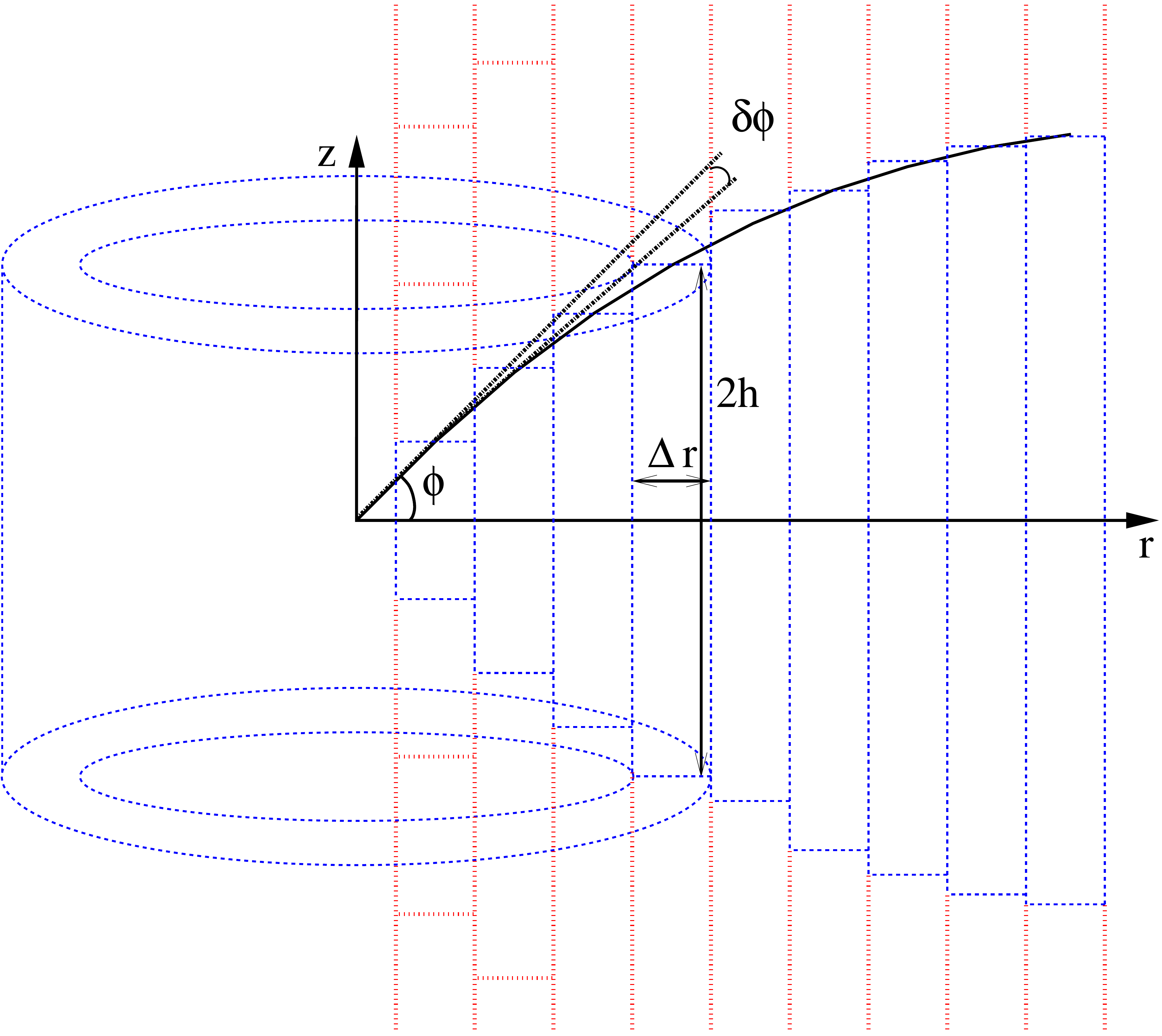} 
\vspace{-0.2cm}
\caption{{\small (Color online) Deflection $\delta \phi$  of a ray (black plain line) by massive rings (blue dashed lines and red dotted lines) of thickness $\Delta r$ 
and height $2h$, symmetric around the $\boldsymbol{z}$ axis and the horizontal $z=0$ plane. 
One ring is represented, as well as the cross-sections of the others (rectangles). 
In addition to the central rings (i.e. of mid-plane at $z=0$, represented with blue dashed lines), 
other rings above $+h$ and below $-h$ (their rectangular cross-sections are shown with red dotted
lines) also contribute to the deflection, although their total effect largely cancels.
The step $\Delta r$ is small compared to the galactic characteristic radial scale  $h_R$. 
The height $h$ is determined at each step by the bent ray.
\label{fig:rings}}}
\end{figure}

\begin{figure*}[tb]
\center
\includegraphics[width=0.8\textwidth]{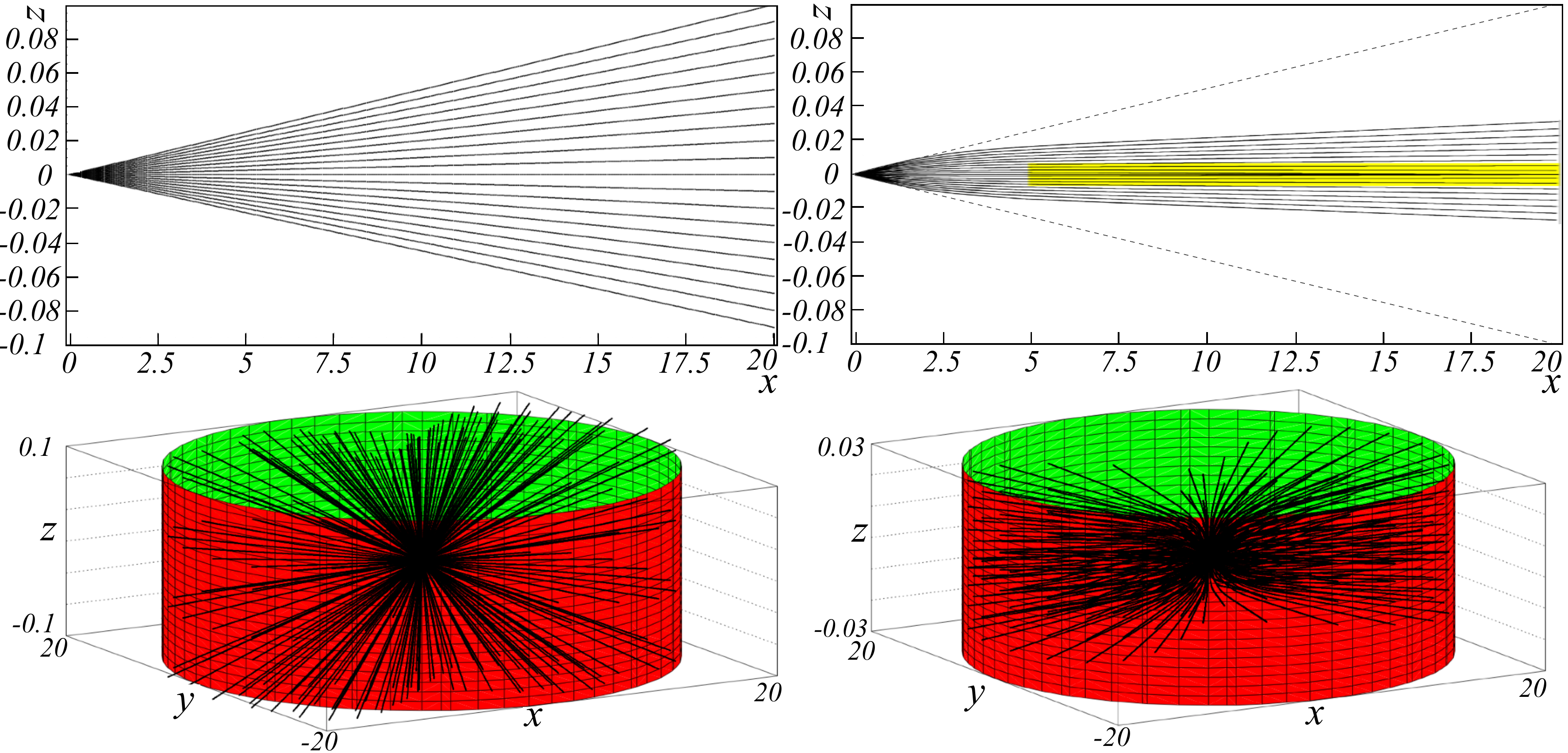}
\vspace{-0.4cm}
\caption{{\small Field lines for a disk centered at $x=y=z=0$, symmetric about $\boldsymbol{z}$, and of 
density decreasing exponentially with radius $r$. 
Left: without field self-interaction. Right: with self-interaction. 
For clarity, only field lines emerging with $| \phi | < 0.005^\circ$ from the center of the disk
are drawn, and the vertical scale on the bottom right panel is zoomed in.
Since densities typical of galaxies cause a small bending, larger densities were used here to make
the effect clearly visible. The figure showing an academic case rather than a physical one, the axis units are therefore unspecified. 
The top panels show the lines in the radial plane ($x,y=0,z$). 
The dashed lines on the right top plot are straight lines with $| \phi | = 0.005^\circ$, i.e. the envelope of the generated lines
when self-interaction is not accounted for. 
The yellow band shows the area where the lines in the ($x,y=0,z$) plane 
are approximately parallel, yielding a $1/r$ behavior of the force. 
\label{fig:lensing_visualization}}}
\end{figure*}

We  will compute numerically the distortion of the field lines. These originate from the galaxy center at angle $\phi$ 
with respect to the $z=0$ plane. 
A small angular deviation $\delta \phi$ of a field line passing near a point mass $\mathcal{M}$ is approximately given by
\begin{equation}
\delta \phi = { {4G\mathcal{M}} \over {hc^2} },
\label{eq:lensing_basic}
\end{equation} 
where $h$ is the impact parameter and $c$ the speed of light. 
For our calculation, we model the galaxy disk as made of concentric rings of radius $r$,  thickness $\Delta r$ and 
height $2h$. The ring masses $\mathcal{M}$ bend the field lines according to Eq.~(\ref{eq:lensing_basic}). 
Fig.~\ref{fig:rings} shows a sketch of the rings and the bending of a field line.
In addition, rings at altitudes $- (2j+1)h$ ($j \in \mathbb{N} $) 
contribute to deflecting more the field lines toward the $z=0$ plane, while
rings at $+ (2j+1)h$ deflect the field lines away from the  $z=0$ plane. 
Nevertheless, the dominant bending comes from the rings with mid-planes at $z=0$, henceforth referred to as ``central rings''. 
For disk galaxies, a good general model for the baryonic matter density is that it exponentially decreases with radius $r$ and altitude $z$ 
according the characteristic scales $h_R$ and $h_z$, respectively~\cite{the:galaxy texbooks}. 
Thus, the mass $\mathcal{M}^j_k$ of a ring of radius $r \equiv k\Delta r$ ($k \in \mathbb{N} $) and of horizontal mid-height plane at altitude $z=jh(r)$ is~\cite{deprojection note} 
\begin{align}
  \mathcal{M}^j_k &  =  { M^*_\mathrm{tot} \over {h_R^2 h_z}}  \int _{(2j-1)h}^{(2j+1)h} \int _{k\Delta r/2}^{(k+1)\Delta r/2}  r {\rm{e}}^{-r/h_R}{\rm{e}}^{-|z|/h_z} {\rm{d}}z{\rm{d}}r, 
\end{align}
with $M^*_\mathrm{tot}$ the total mass of the galaxy (the  label $*$ signals that only baryonic mass is considered). 
Therefore, the mass of a central ring ($j=0$) of radius $ k\Delta r$ is, after integrating over $-h \leq z \leq h$:
\begin{align}
 \mathcal{M}_k  =  {2M^*_\mathrm{tot} \over {h_R^2 }}  \left(  1-{\rm{e}}^{-|z|/h_z}  \right)  \int _{k\Delta r/2}^{(k+1)\Delta r/2}  r {\rm{e}}^{-r/h_R} {\rm{d}}r  ,
\label{eq:M_central_ring}
\end{align} 
and similarly upon integrating over $k\Delta r/2 \leq r \leq (k+1)\Delta r/2$, we obtain:
\begin{align}
 \mathcal{M}_k  =  2M^*_\mathrm{tot}  \left(  1-{\rm{e}}^{-|z|/h_z}  \right)  \left( (a_{-}+1 ){\rm{e}}^{-a_{-}}  -  (a_{+}+1 ){\rm{e}}^{-a_{+}}\right)  ,
\label{eq:M_central_ring}
\end{align} 
with $a_{\pm} \equiv (2k\pm1)\Delta r/(2h_R)$. For non-central rings $j\neq 0$ (which contribute less to the bending than $j=0$ rings),
the d$z$ integration yields  $(  {\rm{e}}^{-(2j-1)|z|/h_z}  -   {\rm{e}}^{-(2j+1)|z|/h_z} )$ rather than $(  1-{\rm{e}}^{-|z|/h_z})$. Therefore, their mass is
\begin{align}
  \mathcal{M}^j_k =  { \mathcal{M}_k \over {2(1-{\rm{e}}^{-|z|/h_z})}}  \left(  {\rm{e}}^{-(2j-1)|z|/h_z}  -   {\rm{e}}^{-(2j+1)|z|/h_z}  \right).
\label{eq:M_ring}
\end{align} 
The angular bending from a ring is then given by:

\begin{equation}
\delta \phi (r,z) = { {G\mathcal{M}^j_k} \over {\pi c^2}}E(r,z) ,
\label{eq:lensing_master}
\end{equation} 
where $E(r,z) \equiv 2{\int _0^\pi \big[\big(2r \sin({\psi \over 2}) \big)^2 +z^2\big]^{-\sfrac{1}{2}} {\rm d} \psi}$ is the complete elliptical integral of the first kind.
The case described in this article is that of pure disk galaxies i.e. Hubble types 5 or 6. 
Using the method for earlier Hubble types, with prominent bulges, would require modifying the mass density 
distribution but does not fundamentally change the method. An example of a calculation is visualized with raytracing 
in Fig.~\ref{fig:lensing_visualization}. For this example, we used densities larger than those typical of galaxies to make the bending 
 of the field lines conspicuous. The bending for actual galaxies densities is small but, as explained next, the ensuing 
effect is magnified at large distances, making its consequence on galactic rotation sizable  at large distances.   
Fig.~\ref{fig:lensing_visualization} elucidates two important facts: 

\hspace{-0.2cm}(1) Although the mass density, and thus the force, 
is larger at small radii, the difference between the field
lines in the right panel and those in the left panel increases with $r$. Hence, 
even if the force is strongest at small $r$, the consequence of self-interaction is
more evident at larger $r$. This explains why the missing mass discrepancy is largest at larger $r$ 
where fields are weak and, seemingly paradoxically, local self-interaction is not important ($\delta \phi \approx  0$).
For example, the field lines for $r \gtrapprox$ 4 
are straight, i.e. $\delta \phi \approx 0$, but are also roughly parallel to each other in a radial plane and for small $z$. This implies  
a force with a $r$-dependence differing from $1/r^2$. This difference would generate a large missing/dark mass in a Newtonian analysis.
In contrast to this straightforward consequence and to empirical observations,
dark matter halo models tend to predict large densities of dark matter at small $r$, which is known as the core-cusp problem~\cite{Core-Cusp Problem}.
Furthermore, this explains why the gravity modifications of MOND~\cite{Milgrom:1983ca}
are enabled below a small characteristic acceleration. 

\hspace{-0.2cm}(2) The field lines at small $r$ display an approximately isotropic distribution, leading to the familiar $1/r^2$-dependence of gravity. 
At larger $r$ and small $z$,  indicated in Fig.~\ref{fig:lensing_visualization} by a yellow band, 
the field lines, while still axisymmetric around $ \boldsymbol{z}$, 
tend to become parallel within radial planes. This leads to a  $\sim 1/r$ behavior of the force and a logarithmic potential $\ln (r)$
within a disk of height about 0.01 in the example of Fig~\ref{fig:lensing_visualization}. For this thin disk, the rotation curves are flat. 
The field lines outside the yellow region show that gravity still acts outside the galaxy,
albeit with depleted strength. This contrasts with the calculations of 
Refs.~\cite{Deur:2009ya, Deur:2016bwq} in which the field needed to be
set to zero outside the disk.

The gravitational force in the radial direction, $F_r(r,z)$, is computed as the flux $\mathcal{F}(r,z)$ of the gravity field through a small surface at radius $r$ and altitude $z$. 
In practice in the numerical calculation, $\mathcal{F}(r,z)$ is given by the number of generated field lines passing through the small surface.
The total number of generated field lines is chosen so that the small fraction of field lines passing through the surface is large enough to make discretization effects,
{\it viz} the fact that $\mathcal{F}(r,z)$ is an integer (number of lines), negligible; see Appendix B.
We verified that in the case without self-interaction ($\delta \phi \equiv 0$), the expected $1/r^2$ dependence of the force is recovered. 
For an idealized galaxy with homogeneous mass distribution, the two-dimensional flux in a horizontal ($z$=constant) plane, $\mathcal{F}_\mathrm{h}$, 
must be unaffected by lensing due to the $ \boldsymbol{z}$-axisymmetry: in a $z$=constant plane, the 
bending from masses on one side of the field line is canceled by the bending from the masses on the 
other side. After verifying numerically that indeed $\mathcal{F}_\mathrm{h} \propto 1/r$, we greatly  
accelerated and simplified the numerical calculation by
generating field lines only in the vertical plane ($x,y=0,z$) and recording the flux $\mathcal{F}_\mathrm{v}$ 
through  a small vertical segment. Then, the radial component of the force
is $F_r(r,z) \propto \mathcal{F}_\mathrm{v}\mathcal{F}_\mathrm{h} \propto \mathcal{F}_\mathrm{v}/r$.

In this article, we are chiefly interested in the $r$-dependence of the force and how it affects galactic rotation curves. For the {\it absolute} determination of the 
force, and thus of the rotation speed, the proportionality constant between $F_r(r,z)$ and $\mathcal{F}(r,z)$ can be determined by equating at small $r$  
the relative $F_r(r,z)$ to the absolute Newtonian expectation, since for $r \ll h_R$ the difference between $F_r(r,z)$ and the Newtonian expectation is negligible, see Fig.~\ref{fig:flux_ex}.
It is worth repeating however that while this difference is negligible for $r \ll h_R$ because the effect of the field line distortions has not propagated 
yet, this region is at the origin of the significant difference at larger $r$, see discussion (1) at the beginning of this Section.

\subsection{Galaxy Model and systematic studies \label{Parameter values}}

The physical parameters of the galaxy model and their chosen nominal values are: \label{Choice of galactic parameter values}
(1) the galaxy total mass $M^*_\mathrm{tot}=3 \times 10^{11}M_{\odot}$. This is a typical baryonic mass of a large spiral galaxy;
(2) the radial scale $h_R$. The present radial scale for spiral galaxies is typically $h_R \approx$ (0.5--5) kpc. 
However, $h_R$ increases with time: in the Milky Way, for instance, $h_R$ for old stars
is about twice smaller than that of young stars~\cite{Amores:2017}. 
The initial (smaller) radial scale must be chosen so we used $h_R=1.5$ kpc;
(3) the scale height $h_z$. The stellar scale height of disk galaxies typically follows $h_z^* \approx 0.1 h_R$, while 
the gas scale height $h_z^\mathrm{gas}$ is smaller: $h_z^\mathrm{gas} \approx (0.02$--$0.04)h_R$. 
We used a typical value $h_z \approx 0.03 h_R$ for two reasons:
firstly, gas friction and attraction between the stars at same $r$ but different $z$ tend to 
make the the denser layers at small $z$ to drag the less dense higher-$z$ layers.
Secondly,  high-density gas produces the stars. Hence, the initial star distribution has a smaller $h_z^*$ than the one
observed in mature galaxies, as demonstrated by the smaller $h_z^*$ of young blue stars compared to the $h_z^*$ of older stars.
Although $h_z^*$ increases with time due to dispersion from two-body interactions, angular momentum conservation 
and the smallness of the vertical speed $v_z$  acquired from two-body interaction, $v_z \ll v_\theta$, 
impose that the rotation speed of the stars remains nearly constant. The rotation speed is thus 
largely determined by the force and density well within a small-$z$ disk before star dispersion occurs.
We varied the parameters within the ranges 
$10^{11}M_{\odot}$$\leq M^*_\mathrm{tot} \leq 10^{12}M_{\odot}$, 
0.5~kpc $\leq h_R \leq $ 5~kpc and
0.01~kpc $\leq h_z  \leq$ 0.5~kpc, 
and found that the self-interaction effect mostly depends on $M^*_\mathrm{tot}$ and $h_z$, and less on  $h_R$. 

In actual galaxies, the smooth decrease of the density with $r$ and $h$ is an average dependence. 
Measurements of HI gas density distribution show that it varies by as much as several times the average. 
It can be assumed that the stellar density fluctuates similarly. 
Since averaging is a linear operation, average quantities may not be adequate inputs for non-linear systems. Thus, to
investigate the effect of density fluctuations, we randomly varied each $\mathcal{M}^j_k$ following a Gaussian distribution centered on the
value that $\mathcal{M}^j_k$ would have assuming an exponential
decrease, and of various widths, with tails  truncated symmetrically so that $\mathcal{M}^j_k >0$.
We found that the effect of the fluctuations is negligible, including the dependence on the width.
Beside random density variations, systematic ones are also possible,
e.g. a warp of the galactic disk. We show in Appendix C  
that such warping does not influence significantly the flux.

\section{Results \label{sec:results}}

\begin{figure}[tb]
\center
\includegraphics[width=0.4\textwidth]{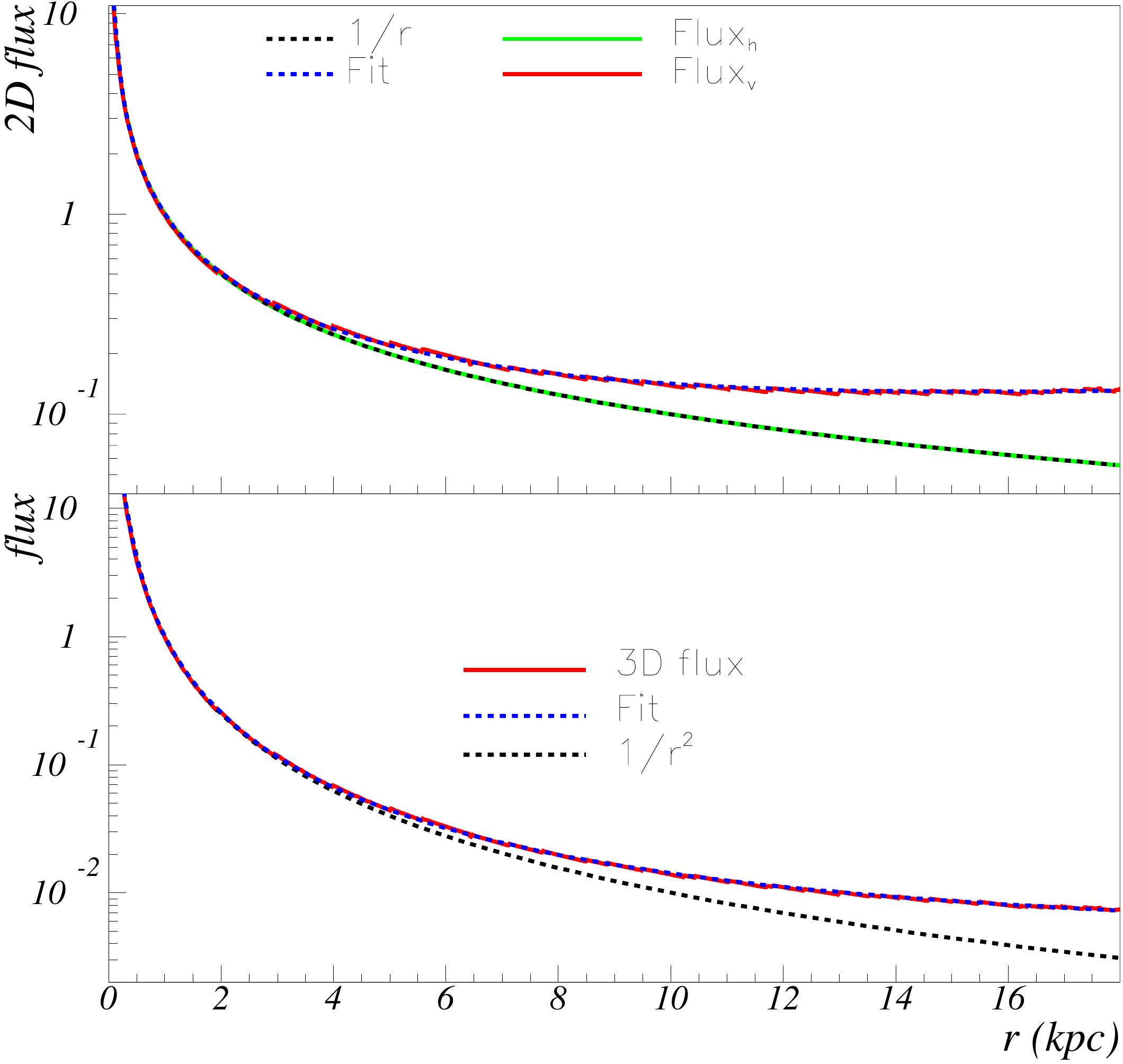} 
\vspace{-0.3cm}
\caption{{\small 
Top: two-dimensional fluxes vs. radial distance $r$.  Green line: horizontal flux $\mathcal{F}_\mathrm{h}(r)$. 
Red line: vertical flux $\mathcal{F}_\mathrm{v}(r)$.  
The modeled galaxy has a total mass  
$M^*_\mathrm{tot} = 3 \times 10^{11}M_{\odot}$, a radial scale length $h_R=1.5$ kpc (deprojection corrections are included), and a scale height $h_z=0.03h_R$. 
The dashed black line is a $1/r$ curve to guide the eye. 
The dotted blue line shows the fit $\mathcal{F}_\mathrm{v} = 1/r+0.0042r$, for $r$ in kpc.
Bottom:  total flux $\mathcal{F}(r)$ (red line) with fit $\mathcal{F}= 1/r^2+0.0042$ for $r$ in kpc (dotted blue line). 
The dashed black line is the Newtonian $1/r^2$ expectation. 
\label{fig:flux_ex}}}
\end{figure}

The  calculation result for the nominal galactic parameters 
is shown in Fig.~\ref{fig:flux_ex} along with fits to the fluxes using the form $\mathcal{F}_\mathrm{v} = 1/r+\alpha r$,
with $[\alpha]=\mbox{kpc}^{-2}$,
and $\mathcal{F} \equiv \mathcal{F}_\mathrm{v}\mathcal{F}_\mathrm{h}= \mathcal{F}_\mathrm{v}/r$. 
Since the force depends on $z$, we averaged it over $h_z$. (For our parameter values, averaging is equivalent to a
calculation done with $M^*_\mathrm{tot}$ reduced by 19\%.)

\subsection{Rotation curves \label{Rotation curves}}
\begin{figure}[tb]
\center
\includegraphics[width=0.4\textwidth]{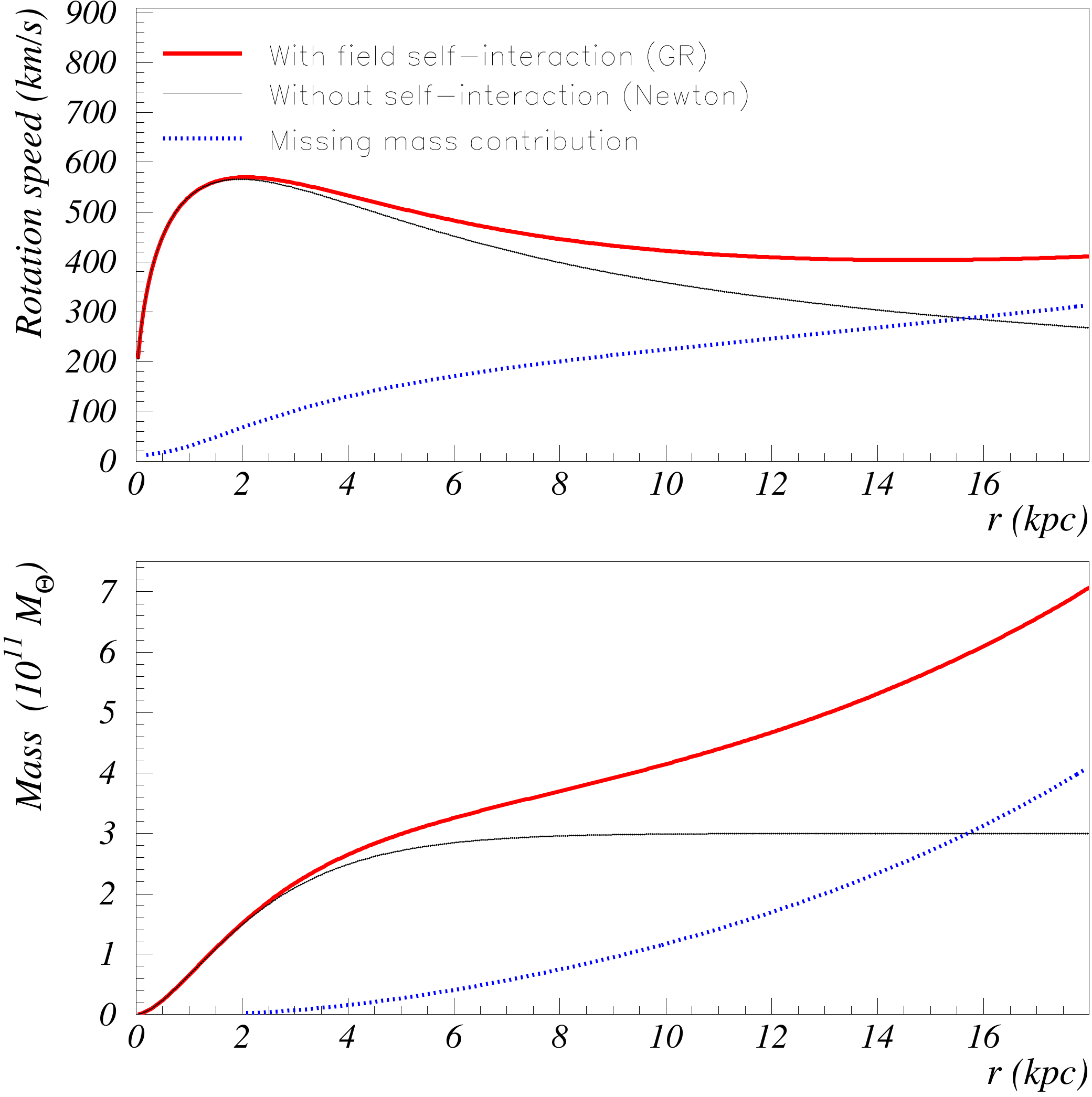} 
\vspace{-0.5cm}
\caption{{\small 
Top: rotation curves generated using the results in Fig.~\ref{fig:flux_ex}. The red line is the rotation
curve accounting for General Relativity's field self-interaction. The black line is without self-interaction (Newtonian case).
The dotted line is the quadratic difference of the two. This represents the missing/dark mass that one would need to introduce 
in an analysis employing Newton's law, rather than General Relativity, in order to recover the red curve. 
Bottom: apparent mass contained within $r$. Each curve corresponds to one of the top panel rotation curves analyzed using Newton's law. 
The red and dotted lines 
are the effective masses obtained from expressing the field self-interaction as an equivalent mass. The black line shows the actual mass.
\label{fig:rot_curves}}}
\end{figure}
Once field self-interaction is effectively included in the calculation of $\boldsymbol{F}(r)$, 
the Newtonian kinematical formalism can be used to 
obtain the rotation curves. 
The radial force is parameterized as 
\begin{equation}
F_r(r) = (Gm_1m_2/r^2)[1+ \alpha r^2], 
\label{Eq:rad force}
\end{equation}
with $\alpha$ determined from the fit to the flux, see Fig.~\ref{fig:flux_ex}. This and the radial equilibrium condition for a body 
in circular orbit at distance $r$ yield the rotation speed: 
\begin{equation}
v_\theta(r) =  \sqrt{GM(r)[1/r+ \alpha r]}. 
\label{Eq:rotcurv}
\end{equation}
Here, $M(r)$ is the mass ,  enclosed within deprojected radius $r$ and determined by using the Abel equation applied to $I(R)\propto {\rm{e}}^{-R/h_R}$, 
with $R$ the projected radius. 
The resulting $v_\theta(r)$ is shown in the top panel of Fig.~\ref{fig:rot_curves} and displays the plateau that is 
typically observed at large $r$. 

We remark that to derive $v_\theta(r)$, the force is approximated as if $M(r)$ were concentrated at $r=0$. While this
is exactly true only for spherically symmetric matter distributions and $F \propto 1/r^2$ (Newton's shell theorem), 
the sharp density peaking near the galaxy center 
makes this approximation acceptable in the present case. To check this,
we performed a $N-$body simulation using both the Newtonian and the perturbative post-Newtonian formalisms and compared
the resulting rotation curves to that obtained using the shell theorem. They are comparable. For the
non-perturbative case, the field lines tend to  become confined within the disk~\cite{Deur:2009ya, Deur:2016bwq} and in that limit, the resulting $1/r$ force makes the (now in two dimensions) shell theorem exactly valid  again.

\subsection{Effective dark mass profile}
The effective dark mass profile can be obtained straightforwardly. In a dark matter framework $F = GMm/r^2$ and 
$v_\theta(r)=[ G(M^*(r)+M_\mathrm{dark}(r))/r]^{\sfrac{1}{2}}$, where 
$M^*(r)$ and $M_\mathrm{dark}(r)$ are the baryonic and dark masses enclosed within $r$, respectively. The equivalent dark mass profile is then:
\begin{equation}
M_\mathrm{dark}(r)=rv^2_\theta(r)/G -M^*(r).
\label{eq:rDM profile}
\end{equation} 
This is shown by the dotted line in the bottom panel of Fig.~\ref{fig:rot_curves}. 
Its rise with $r$ illustrates the earlier discussion 
that the consequence of self-interaction is magnified at larger $r$ despite negligible local self-interactions there.

\section{Predictions and verifications}

\begin{figure}[tb]
\center
\includegraphics[width=0.45\textwidth]{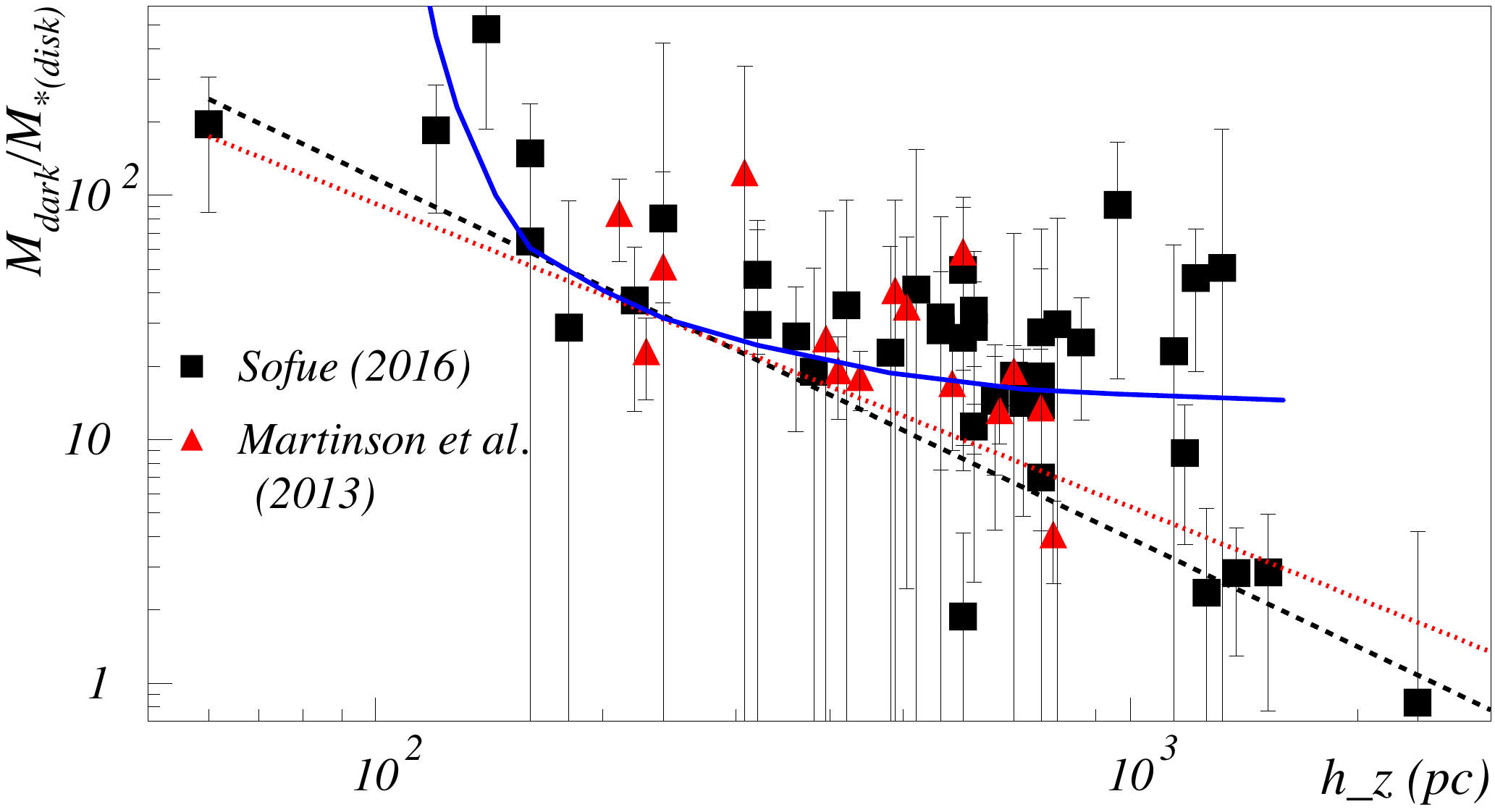} 
\vspace{-0.3cm}
\caption{{\small  Galactic dark mass $ M_\mathrm{dark}$ normalized to baryonic disk mass vs. vertical scale length $h_z$. 
The squares show the data from Ref.~\cite{sofue15} and triangles from \cite{Martinsson:2013ukc}. 
$ M_\mathrm{dark}/M^*_\mathrm{disk}$ correlates with $h_z$. The dashed and dotted lines give the best fit to Refs.~\cite{sofue15} and \cite{Martinsson:2013ukc}, 
respectively, assuming a $ah_z^b$ form. The plain line is the lensing calculation described in this article. 
\label{fig:hz_Mdm1}}}
\end{figure}

The model shows that the effect of field self-interaction depends mainly on the total baryonic mass $M^*_\mathrm{tot}$,
specifically on the galactic density at small $r$,  and the scale height $h_z$. 
The latest suggests that the inferred dark mass $M_\mathrm{dark}$ of a galaxy correlates with its $h_z$.
We check this prediction by using two different sets of data~\cite{sofue15,Martinsson:2013ukc} 
that provide $M_{dark}$ and disk characteristic scales. 
The first set, from Sofue~\cite{sofue15}, provides only the radial scale $h_R$, but the proportionality 
relation, $h_z \simeq \epsilon h_R$, allows us to perform the analysis (a correlation analysis is independent of the value of $\epsilon$). 
The second set, from Martinson~{\it et al.}~\cite{Martinsson:2013ukc}, provides both $h_z$ and $M_{dark}$. 
Fig.~\ref{fig:hz_Mdm1}  shows $M_\mathrm{dark}/M^*_\mathrm{disk}$ vs. $h_z$ for both sets 
(set~\cite{Martinsson:2013ukc} was recombined in $200$ pc  bins of $h_z$ for clarity). 
We normalized  $M_\mathrm{dark}$ by the disk baryonic mass $M^*_\mathrm{disk}$ to cancel the expected dependence 
of GR's non-linearity with the baryonic mass, as well as other known relation between 
$h_R$ ({\it viz} $h_z$) and luminosity or $M^*_\mathrm{disk}$.
The predicted correlation is clearly visible: fitting with $ M_\mathrm{dark}/M^*_\mathrm{disk} = ah_z^b$ yields $b\neq0$;
$b=-1.48 \pm 0.11$ for set~\cite{sofue15} and $-1.25 \pm 0.14$ for~\cite{Martinsson:2013ukc}.
Although the galaxies in the two sets are distinct, the two values of $b$ agree. 
We can also quantify the degree of correlation by calculating the Pearson linear correlation coefficient
between $\ln (M_\mathrm{dark}/M^*_\mathrm{disk})$ and $\ln h_z$. It is $-0.70$ and $-0.69$
for sets~\cite{sofue15} and \cite{Martinsson:2013ukc} respectively. 
Normalizing $M_\mathrm{dark}$ to the total baryonic mass $M^*_\mathrm{disk+bulge}$ yields 
similar results but with lower values of $b$:
$-1.10 \pm 0.13$ and $-1.5 \pm 0.30$ for \cite{sofue15} and \cite{Martinsson:2013ukc}, respectively. 
This is because no causal connection is expected between $h_z$  and $M^*_\mathrm{bulge}$.
Also shown in Fig.~\ref{fig:hz_Mdm1} are the results of our model calculated for 
$M^*_\mathrm{tot}=0.9\times10^{11}$~M$_\odot$ (the approximate average of $M^*$ for set~\cite{sofue15})
and $h_R=1.5$~kpc. 
In Refs.~\cite{sofue15} and \cite{Martinsson:2013ukc}, $M_\mathrm{dark}$
is estimated at $R_{200}$ which ranges from tens  to hundreds of kpc. 
Hence, we calculated our effective dark mass up to $r=100$~kpc. 
The result (not adjusted  to the data) is shown by the plain line in Fig.~\ref{fig:hz_Mdm1}. 
It agrees with the observations, with a  
$\chi^2$ similar to that of the $ah_z^b$ fit.  

\
\section{Summary}

The values of galactic masses and characteristic distances suggest that field self-interaction, a feature of general relativity (GR), 
needs to be included in studies of galaxy dynamics~\cite{Deur:2009ya, Deur:2013baa, Deur:2019kqi}. 
Field self-interaction increases gravity's strength compared to the Newtonian expectation. 
The effect will become noticeable in systems of large enough mass. 
In a Newtonian analysis of such systems, the effect would be misinterpreted as a missing mass (dark matter).
Furthermore, the similarities between, on the one hand the GR's Lagrangian and observations linked to 
dark matter and energy, and on the other the QCD Lagrangian and hadron structure phenomenology, offer another compelling reason 
to investigate GR self-interaction as an explanation for the universe's dark content. 
Finally, the exclusion by direct searches of most of the natural phase space for WIMP and axion candidates, and the 
emergence of dark energy from GR self-interactions~\cite{Deur:2017aas}, make this explanation of the 
missing mass problem more plausible. 

In this article, we presented a new approach to compute the effects of GR's self-interaction based on a mean-field
technique and the formalism of gravitational lensing. 
We find that self-interaction effects are important for galaxy dynamics and tend to flatten rotation curves. This agrees with  
Refs.~\cite{Deur:2009ya, Deur:2016bwq} in which a different 
method (numerical lattice calculation of path integrals) was used to account for field self-interaction.
The present method is faster and applicable to any mass distribution. 
However, it is less directly based on GR's equations than the path integral approach.
Beside flattening the galactic rotation curves, the method explains straightforwardly why the missing mass
discrepancy worsens at large galactic radii.
In contrast, dark matter halo models must be tuned to the specific density profile of a given galaxy to flatten its rotation curve 
(disk-halo conspiracy), and they tend to predict larger dark matter density at small radii (core-cusp problem).
The present approach predicts a correlation between the missing mass inferred in galaxies and their vertical scale length. We verified this
correlation with two separate data sets.

\section*{Acknowledgments}
The author thanks B. Guiot, C. Sargent, S. \v{S}irca, B. Terzi\'c and X. Zheng for useful discussions.

~

\noindent{\bf Appendix A: \small{calculation with fluxes originating from the complete mass distribution}}

\begin{figure} [h]
\centering
\vspace{-0.2cm}
\includegraphics[width=0.4\textwidth]{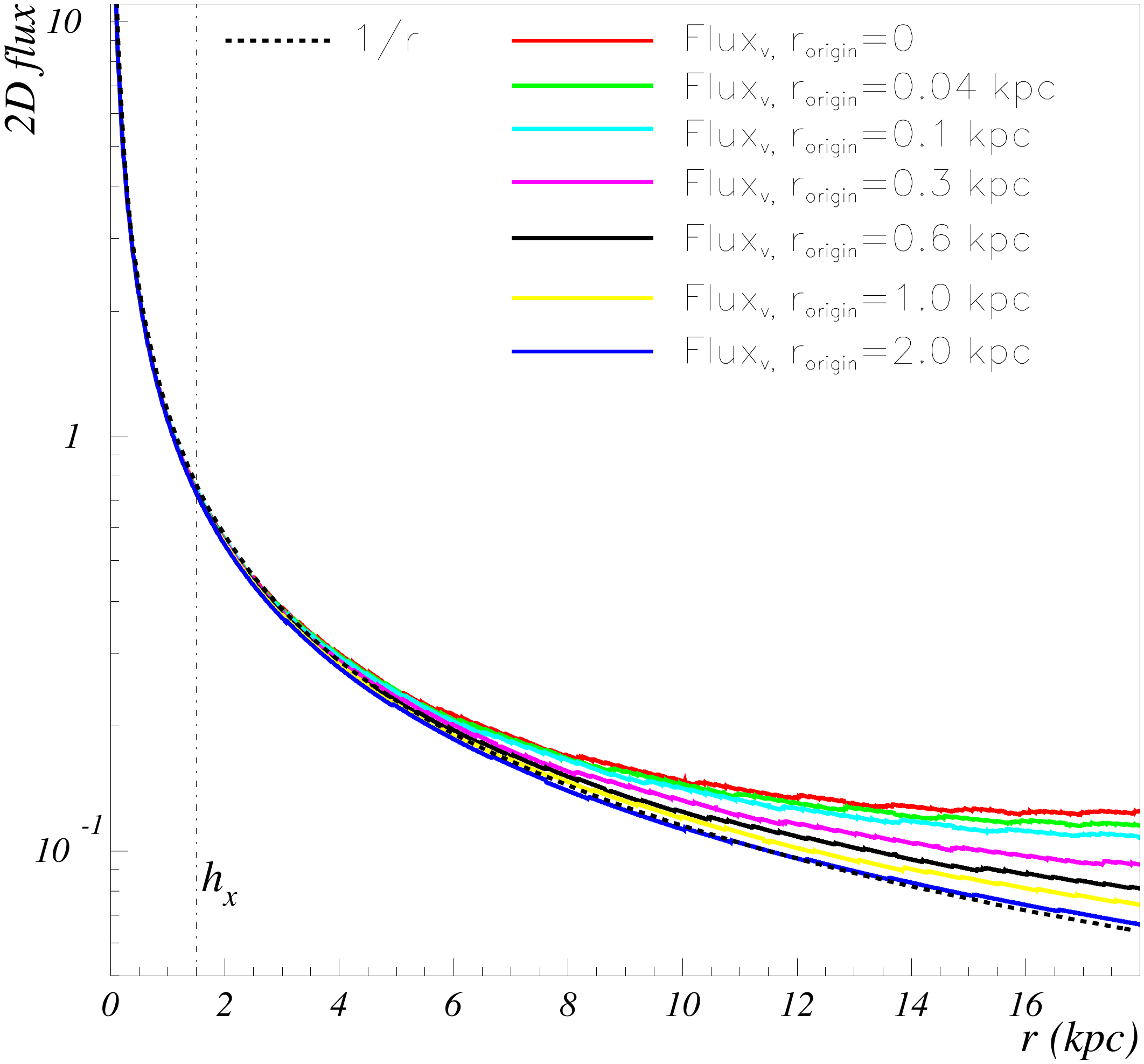}
\vspace{-0.4cm}
\caption{ \label{fig:flux_not_centered}
 {\small  Two-dimensional flux $\mathcal{F}_\mathrm{h}$ vs. galactic radius $r$. $\mathcal{F}_\mathrm{h}$ is
calculated for various points of origin $r_{origin}$ along $r$. 
The result in red is for $\mathcal{F}_\mathrm{h}$ originating from the galaxy center, $r=0$, as depicted in 
Fig.~\ref{fig:lensing_visualization}. The green curve is for $\mathcal{F}_\mathrm{h}$ 
originating at a point at $r=0.04$~kpc. The points of 
origin for the other curves, from 0.1 to 2~kpc,  are given in the figure. 
The results are for the galactic parameters listed in Fig.~\ref{fig:flux_ex}. The radial characteristic scale 
$h_r=1.5$~kpc is shown by the vertical dot-dashed line.
The dashed curved is the Newtonian $1/r$ two-dimensional flux.
The effect of field self-interaction is noticeable for fluxes originating at $r_{origin} \lesssim h_r$. 
To aid the comparison, each flux is normalized to the same
maximum value, and plotted with its $r$-dependence shifted by $r_{origin}$.  }}
\end{figure}

In this appendix, we verify that approximating the flux as emerging from $r=0$, as shown
in Fig.~\ref{fig:lensing_visualization}, is justified for a typical disk galaxy. To do so,
we compare the calculation obtained in the main part of this article (Fig.~\ref{fig:flux_ex})  to 
a calculation accounting for fluxes originating from $r_{origin} \neq0$.
Fig.~\ref{fig:flux_not_centered} shows some of these fluxes. 
We see that field self-interaction distorts the flux only when
it originates from points $r_{origin} \lesssim h_r$ (e.g. red or green curves). 
The figure also shows that the distortion becomes manifest only at $r$ larger than several $h_r$. 

\begin{figure}
\centering
\includegraphics[width=0.4\textwidth]{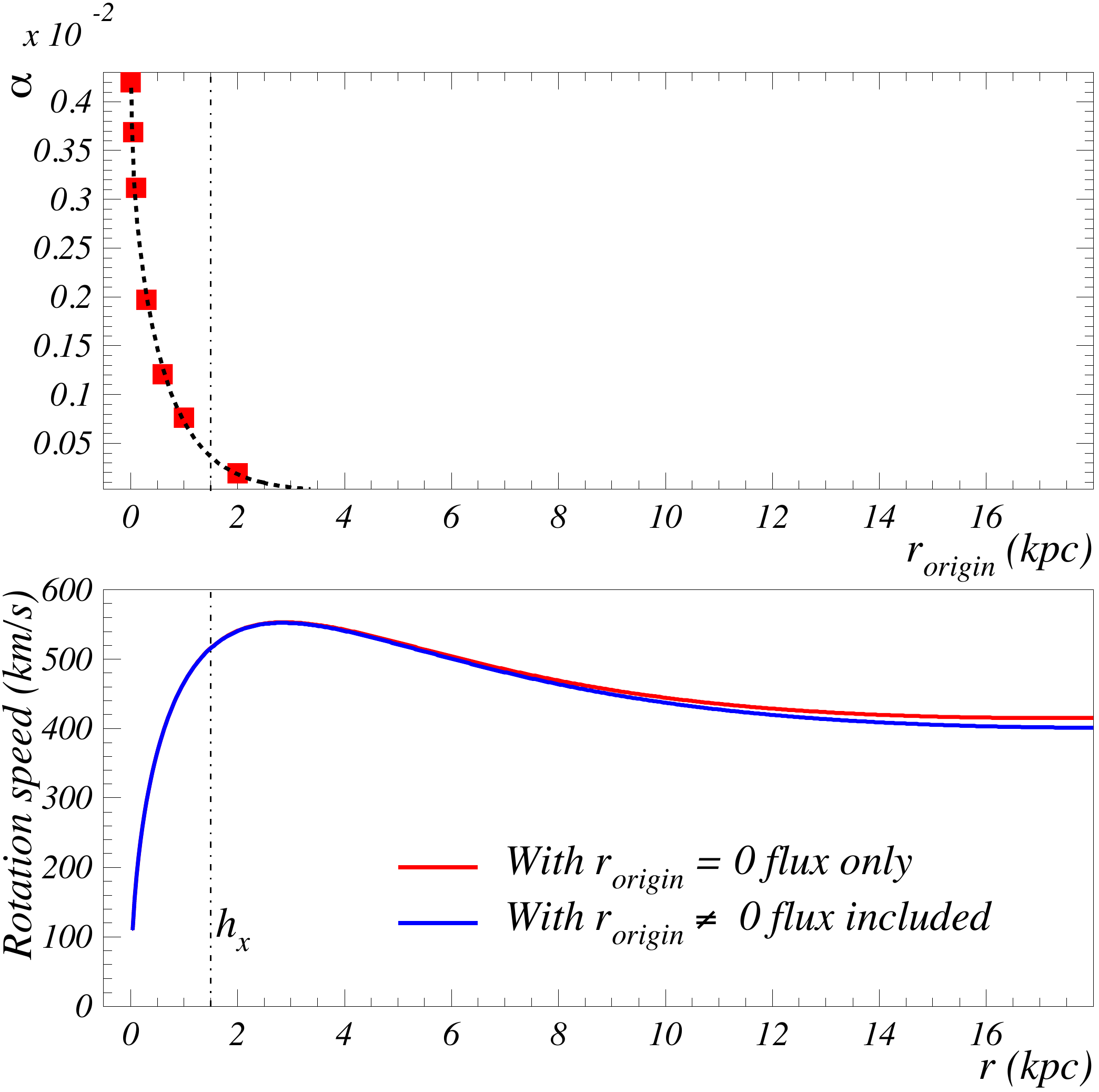}
\vspace{-0.4cm}

\caption{\label{fig:non_cent_flux_cor}
{\small 
 Top panel: The parameter $\alpha$ characterizing the non-Newtonian flux behavior, vs. $r_{origin}$.  
Bottom panel: Galactic rotation curve accounting for the fact that fluxes originate at different $r_{origin}$ (blue line). The red line shows the result if the full flux is approximated to originate at $r=0$.  }}
\end{figure}

The next step is to show that the flux in Fig.~\ref{fig:flux_ex} 
is representative of the average flux, i.e. the more distorted fluxes from  $r_{origin} \ll h_r$ and the
less distorted fluxes from $r_{origin} \gtrsim h_r$ approximately average to the flux in Fig.~\ref{fig:flux_ex}.
The top panel  of Fig.~\ref{fig:non_cent_flux_cor} shows the evolution with $r_{origin}$ of the parameter 
$\alpha$ defined in Section~\ref{sec:results} and characterizing the flux distortion.  
The squares show $\alpha$ determined from fitting the curves in 
Fig.~\ref{fig:flux_not_centered}. 
The steep behavior of $\alpha(r_{origin})$ means that the structure of the galaxy center 
must be accounted for if the $r_{origin}=0$ approximation is not used. 
This is because even though  $\alpha$ is large only below a few $0.1h_r$, the enclosed mass 
is only a few \% of the total galaxy mass and therefore, the averaged contribution from fluxes with 
$\alpha \neq 0$ is largely diluted by those with $\alpha \approx 0$. However,  an  
additional mass at $r\approx 0$ due to a galactic nuclear cluster  
magnifies the largest value of $\alpha$, $\alpha(r_{origin}\approx 0)$, and although the nuclear cluster 
mass remains negligible compared to the total galaxy mass, this is enough to balance 
the dilution. 
The blue line in the bottom panel  of Fig.~\ref{fig:non_cent_flux_cor} shows the rotation curve 
obtained when the $r_{origin}= 0$ approximation is lifted. 
The red line shows the result obtained with the $r_{origin}= 0$ approximation. 
In both cases, the $3\times 10^{11}$~$M_\odot$ galaxy model includes a $10^{8}$~$M_\odot$ nuclear cluster 
of 15 pc characteristic scale. 
The two curves are close to each others, revealing that averaging the $r_{origin} \neq 0$ fluxes 
or approximating all the fluxes as stemming from $r_{origin} = 0$ happens to lead to similar results for
a typical galaxy mass and typical size and mass of the galaxy cluster. 
The critical role played by the central mass, once $r_{origin} \neq 0$ fluxes
are considered, and the fact that the galactic core and  central black hole masses
correlate~\cite{Ferrarese:2000se}, may explain the clear correlation seen between 
the amount of the missing mass in a galaxy and the mass of its central 
black hole~\cite{Ferrarese:2002ct}.

~

\noindent{\bf Appendix B: \small{Numerical calculation and optimization\label{appendix B}}}

The self-interaction effect is calculated numerically. 
The calculation parameters, along with the checks done to verify that the results are independent of these parameters, are:
(1) the grid radial step $d_\mathrm{step}$, with $d_\mathrm{step} \ll h_R$ to avoid discretization artifacts. 
Varying it between $0.0125h_R \leq d_\mathrm{step} \leq 0.05h_R$ yields no appreciable effect on the results;
(2) the segment length $l_{\mathcal{F}_\mathrm{v}}$ through which the flux $\mathcal{F}_\mathrm{v}$ is calculated. Varying it over $10^{-15}h_z \leq l_{\mathcal{F}_\mathrm{v}} \leq 10^{-2}h_z$ 
shows that  $\mathcal{F}_\mathrm{v}$ plateaus for $l_{\mathcal{F}_\mathrm{v}} \lesssim 10^{-10}h_z$;
(3) the average altitude $z_\mathrm{center}$ and initial transverse size $h_\mathrm{max}$ along which the set of parallel field lines is generated. 
They are generated at altitude $z_\mathrm{center}\pm h_\mathrm{max}$ with $h_\mathrm{max} \gg l_{\mathcal{F}_\mathrm{v}}$ to avoid discretization artifacts. 
Varying $h_\mathrm{max}$ between $100l_{\mathcal{F}_\mathrm{v}} \leq h_\mathrm{max} \leq 200l_{\mathcal{F}_\mathrm{v}}$ yields no change of the results;
Nominally, field lines originate from the mid-plane of the galaxy, $z_\mathrm{center}=0$; 
(4) the number of field lines generated, $n_\mathrm{fl}$. Varying $n_\mathrm{fl}$ between $10^3 \leq n_\mathrm{fl} \leq 10^4$ yields identical results except for expected
discretization effects;
(5) the number of vertical discretization steps $J_\mathrm{max}$, i.e. the number of rings stacked up at same $r$, see Fig.~\ref{fig:rings}. 
Varying $J_\mathrm{max}$ over $1 \leq J_\mathrm{max} \leq 60$ shows that $\mathcal{F}_\mathrm{v}$ plateaus for $J_\mathrm{max} \gtrapprox 50$.

A problem reducing the calculation efficiency is the need to focus on the flux in the small-$z$ disk while 
most of the field lines -- generated from the galaxy center and within a chosen range of $\phi$ -- exit this volume 
since they are only moderately bent. For example, only a few of the lines generated remain in the 
yellow band of Fig.~\ref{fig:lensing_visualization}  (top right panel).
Furthermore, the force must be calculated as $\mathcal{F}_\mathrm{v}$ through a vanishingly small segment, to avoid 
bias by averaging over steeply varying quantities (namely the $z$-dependence of the force).
These problems cannot be solved by choosing a small range for $\phi$, e.g.
(small-$z$ disk height)/($r_\mathrm{max}$) with $r_\mathrm{max}$ the galaxy radial size, because the segment through which $\mathcal{F}_\mathrm{v}$ is calculated
would not be fully covered at small $r$, thus yielding an artificially constant force.
To solve these problems, instead of radial field lines, we generate a set of parallel lines, each with $\phi(r=0)=0$ and drawn 
uniformly around an altitude range $z_\mathrm{center} \pm h_\mathrm{max}$. We then compute the lensing effect on that set 
and weigh the result by $1/r^2$. This allows us to compute $\mathcal{F}_\mathrm{v}$ through an arbitrarily 
small segment at a given $z$ and over the full $r$ range, thereby obtaining the force $\boldsymbol{F}(r,z)$. 
One may object that this method would bias the calculation since for radial lines, 
$h$ in Eq.~(\ref{eq:lensing_basic}) increases with $r$ (providing that lensing is moderate so that
$\phi(r)$ remains positive), while for initially parallel lines, $h$ is constant (if lensing is negligible) or decreases. This would 
lead to underestimate of both the distance of closest approach and $\mathcal{M}$. If the matter density is constant,
the two effects exactly cancel each other in  Eq.~(\ref{eq:lensing_basic}) and there is no calculation bias. 
With $h_\mathrm{max} \ll h_z$ the density is indeed essentially constant ($h_\mathrm{max}=10^{-15}h_z$, see Section~\ref{Parameter values}).
Furthermore, even if the density varies within $h_\mathrm{max}$, a bias would still be insignificant since one also has
$h_\mathrm{max} \ll z_\mathrm{center}$: with $h_\mathrm{max} \ll h_z$ and $h_\mathrm{max} \ll z_\mathrm{center}$, the increase of $h$ with $r$ for a radial line
is negligible.
In fact, we explicitly checked that the two methods of generating radial or parallel lines agree when $h_\mathrm{max} \ll h_z$.

~

\begin{figure}[h]
\centering
\includegraphics[width=0.4\textwidth]{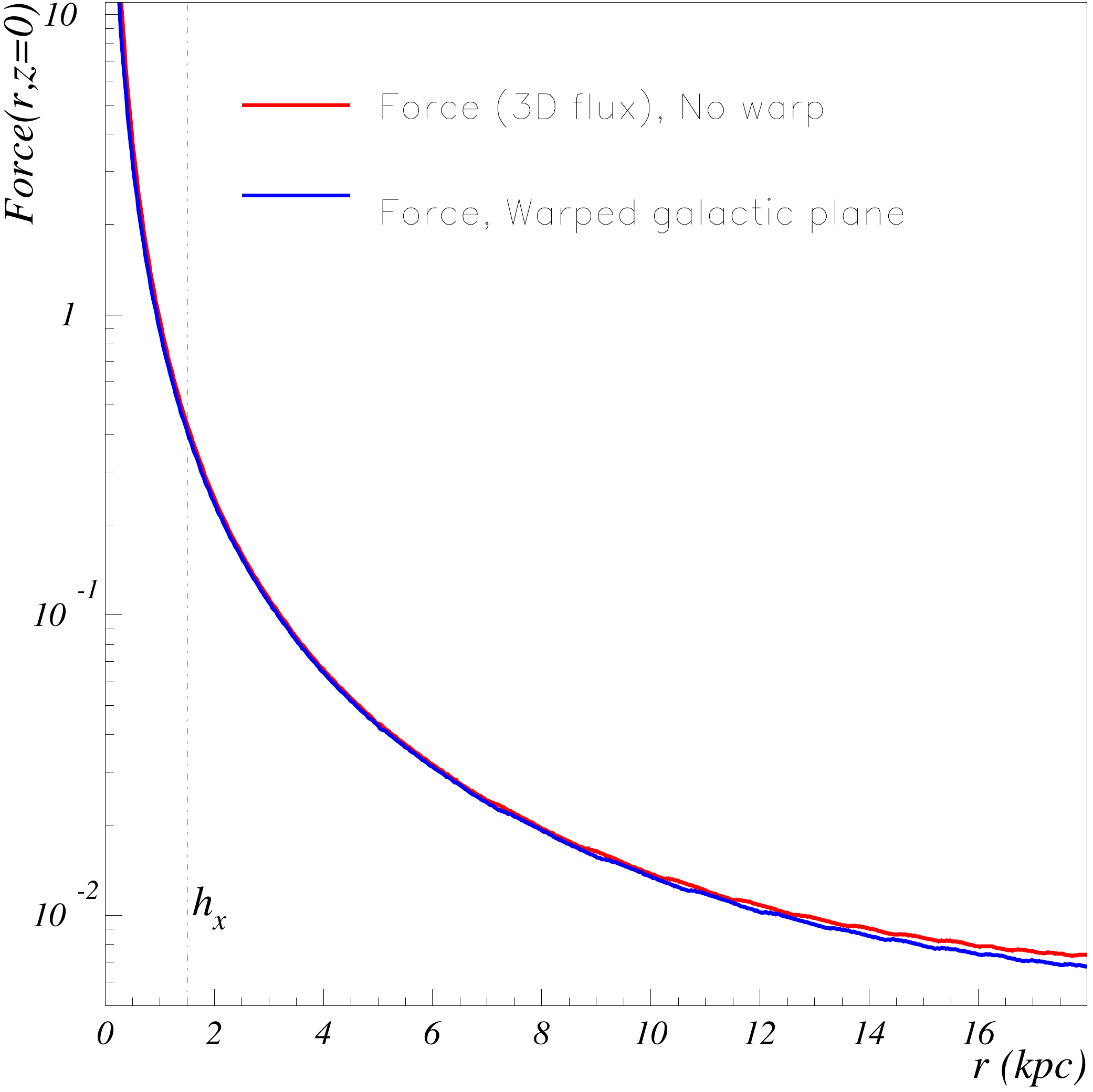}
\vspace{-0.4cm}
\caption{\label{fig:warp}
 In-plane ($z=0$) flux ({\it viz} force in arbitrary unit) in function of radius $r$ for a flat (red curve) and warped 
 disk (blue curve). The curves are calculated with the galactic parameters given in Fig.~\ref{fig:flux_ex}. 
 The radial characteristic scale $h_r=1.5$~kpc is indicated by the vertical dot-dashed line.}
\end{figure}

\noindent{\bf Appendix C: Effect of galactic disk warping \label{subsec:warp}}
Disk warping is often present in disk galaxies. Here, we check how it affects the flux.  
We modeled the disk as having a maximum departure of $2h_z$ from the flat plane.
The departure grows exponentially with a characteristic scale of $3h_r$, being largest 
at the edge of the galaxy. 
Although this is a substantial deformation, its effect on the potential 
 is small, see Fig.~\ref{fig:warp}. This is expected, since  by definition the deformation
 becomes important at large $r$ and there, the mass distribution has little influence
 on the flux, as we showed earlier and discussed in details in Appendix A. 
 We note that the flux is  computed in the $z=0$ plane. 
 Since the mass distribution is in average at $z\neq0$, the force at $z=0$ 
 does not determine the rotation curve. Providing it demands to include the cause of the warp which may
 be external to the galaxy dynamics (e.g. caused by satellite galaxies) and thus lays beyond the scope of this article.


\begin{thebibliography}{00}
\bibitem{deprojection note} For simplicity, Eqs.~(\ref{eq:M_ring})-(\ref{eq:lensing_master}) are derived with a mass density decreasing exponentially 
with deprojected radius $r$. However, the final results are obtained with the proper density derived 
with the Abel integral equation applied to $I(R,z)$, with R the projected radius.
\bibitem{apAnote} The field lines distortion is calculated 
using the actual mass distribution. If all masses are 
taken to be at $r=0$ there is no field line distortion.
\end{thebibliography}
\end{document}